\documentclass[aps, showpacs,showkeys,pra,twocolumn] {revtex4}
\usepackage{hyperref}
\usepackage{amsmath }
\usepackage{amssymb}
\usepackage{epsfig}
\usepackage{natbib}
\newcommand*{\be}{\begin{equation}}
\newcommand*{\ee}{\end{equation}}

\begin{document}
\bibliographystyle{revtex}

\title{\bf Rashba spin-orbit interaction in  quantum  \\ring  with confining potential
of finite depth }
\author{V. V. Kudryashov}
\email{kudryash@dragon.bas-net.by}

\affiliation{Institute of Physics, National Academy of Sciences of
Belarus, \\ 68 Nezavisimosti  Ave., 220072, Minsk,  Belarus}

\begin{abstract} We present the exact wave functions and energy
levels for an electron in a two-dimensional quantum ring with
confining potential of finite depth in the presence of the Rashba
spin-orbit interaction.
\end{abstract}

 \maketitle

\section{Introduction}

Experimental and theoretical investigations of  the spin-orbit
coupling in semiconductor nanostructures have attached
considerable attention.

The Rashba spin-orbit interaction in law-dimensional semiconductor
structures has been the object of many investigations in recent
years. The Rashba interaction is of the form \cite{ras,byc}
\begin{equation}
V_R= \beta_R (\sigma_x p_y - \sigma_y p_x)
\end{equation}
with standard Pauli spin-matrices $\sigma_x$ and $\sigma_y$. The
Rashba interaction can be strong in semiconductor heterostructures
and its strength can be controlled by an external electric field.

Recently, advanced growth techniques have made it possible to
fabricate high quality semiconductor rings. The circular quantum
rings (nanorings) in semiconductors can be described as
effectively two-dimensional systems in confining potential
$V_c(\rho)$  ($\rho =\sqrt{x^2 +y^2}$). The parabolic and infinite
hard wall confining potentials are the most  commonly used
approximations for the quantum rings \cite{song}. As it was
noticed in \cite{ban}, these models do not allow us to study the
tunneling effects and do not permit the existence of unbounded
states. In papers \cite{ban, gro}, a  more realistic model which
corresponds to a quantum ring with a potential well of finite
depth $V = constant$ was proposed:
\begin{equation}
V_c(\rho) =  \left\{
\begin{array}{cl}
 V, & 0<\rho < \rho_i ,\\
 0, &  \rho_i < \rho <  \rho_o ,\\
 V, &  \rho_o < \rho < \infty ,
 \end{array} \right.
 \end{equation}
 where $\rho_i$ and $\rho_o$ are the inner and outer radii
of the ring, respectively. However, this potential was not used in
the presence of the Rashba spin-orbit interaction.
 For example, in paper \cite{shen}, the Rashba interaction was considered for the
infinite hard wall. In this model $V = \infty$.

In the present  paper, we examine a model which corresponds to the
Rashba interaction (1) and the confining potential (2). We shall
present the exact coordinate wave functions and energy levels for
this model.

\section{Analytical solutions of the  Schr\"odinger equation}

 The single-electron wave functions satisfy the
Schr\"odinger equation
\begin{equation}
 \left(\frac{{\bf p}^2}{2 \mu} + V_{c}(\rho) +V_R\right) \Psi = E
 \Psi ,
\end{equation}
where $\mu$ is the effective electron mass.

 The Schr\"odinger equation
 is considered  in the cylindrical coordinates
 $x = \rho \cos \varphi,  y = \rho \sin \varphi$.
 Further it
is convenient to employ dimensionless  quantities
 \begin{eqnarray}
 e &=&\frac{2 \mu}{\hbar^2} \rho_o^2 E, \quad
  v= \frac{2 \mu}{\hbar^2} \rho_o^2 V , \quad
 \beta=\frac{2 \mu}{\hbar} \rho_o \beta_R , \nonumber \\
  r&= &\frac{\rho}{
 \rho_o}, \quad r_i = \frac{\rho_i}{ \rho_o}.
 \end{eqnarray}
 Note that $1 - r_i = (\rho_o - \rho_i)/\rho_o$  is a relative width of a
 ring.

As it was shown in \cite{bul,tsi}, Eq. (3) permits the separation
of variables
\begin{eqnarray}
 \Psi_m(r,\varphi)& =&  u(r) e^{i m \varphi}\left(\begin{array}{c}
 1 \\
  0
 \end{array}\right) + w(r)  e^{i (m+1) \varphi} \left(\begin{array}{c}
 0 \\
  1
 \end{array}\right) , \\
  m&=&0, \pm 1, \pm 2, \ldots \nonumber
\end{eqnarray}
due to conservation of the total angular momentum
$L_z + \frac{\hbar}{2} \sigma_z $.

 In the examined model, we look for the radial wave functions
$u(r)$ and $w(r)$ regular at  the origin $r=0$ and decreasing at
infinity $r \rightarrow \infty$.

We consider three regions $ 0 <r < r_i$ (region 1), $r_i <r <1$
(region 2) and $ 1 <r < \infty$ (region 3) separately.

In the regions 1 and 3 $(v>0)$ we have the following radial
equations
\begin{eqnarray}
r^2 \frac{d^2u}{dr^2} &+& r \frac{d u}{d r}  -(k^+_o k^-_o r^2 +
m^2)u  \nonumber \\
&= &- i ( k^+_o -  k^-_o) r^2\left(\frac{d w}{d r} +
\frac{m+1}{r}w
\right), \nonumber \\
 r^2\frac{d^2w}{dr^2} &+& r \frac{d w}{d r}  -(k^+_o k^-_o r^2 +(
 m+1)^2)w
 \nonumber \\
&= &i( k^+_o -  k^-_0)  r^2 \left(\frac{d u}{d r}- \frac{m}{r} u
\right) ,
\end{eqnarray}
 where
\begin{equation}
k^{\pm}_o(e,v,\beta) = \sqrt{v-e -\frac{\beta^2}{4}} \pm i
\frac{\beta}{2}.
\end{equation}
Note that in the case $\beta = 0$ ($k^+_o  = k^-_o$)  Eqs. (6) are
the modified Bessel equations. In the regions 1 and 2 we must
select the different particular solutions of Eqs. (6) in order to
reproduce the correct behavior of the wave functions at $r \to 0$
and at $r \to \infty$.

In the region 1 using the known properties \cite{abr}
\begin{eqnarray}
 \left(\frac{d}{d r} + \frac{n}{r} \right) I_n(k r)
 &=& k I_{n-1}(k r) , \nonumber \\
 \left(\frac{d}{d r} - \frac{n}{r} \right) I_n(k r)
 &=& k I_{n+1}(k r)
\end{eqnarray}
of the modified Bessel functions  $I_n(k r)$
 it is easily to get the exact solutions
 \begin{eqnarray}
u_1(r) &= &c_1 f_1(m,r) + d_1 g_1(m,r), \nonumber \\
 w_1(r)& = & -c_1 g_1(m+1,r) + d_1 f_1(m+1,r)
\end{eqnarray}
 of system (6) where
 \begin{eqnarray}
 f_1(m,r) = \frac{1}{2}\left(I_m(k^-_o r) + I_m(k^+_o r) \right),
 \nonumber \\
g_1(m,r) = \frac{i}{2}\left(I_m(k^-_o r) - I_m(k^+_o r) \right)
\end{eqnarray}
are the real linear combinations of the modified Bessel functions
with complex arguments. Here $c_1$ and $d_1$ are arbitrary
coefficients. The radial wave functions $u_1(r)$ and $w_1(r)$ have
the desirable behavior at the origin.

In the region 3 using the known properties \cite{abr}
\begin{eqnarray}
 \left(\frac{d}{d r} + \frac{n}{r} \right) K_n(k r)
 &=& -k K_{n-1}(k r) , \nonumber \\
 \left(\frac{d}{d r} - \frac{n}{r} \right) K_n(k r)
 &= &-k K_{n+1}(k r)
\end{eqnarray}
of the modified Bessel functions  $K_n(k r)$
 it is simply to derive the exact solutions
\begin{eqnarray}
u_3(r)& = &c_3 f_3(m,r) + d_3 g_3(m,r), \nonumber \\
 w_3(r)& = & c_3 g_3(m+1,r) - d_3 f_3(m+1,r)
\end{eqnarray}
 of system (6) where
 \begin{eqnarray}
 f_3(m,r) = \frac{1}{2}\left(K_m(k^-_o r) + K_m(k^+_o r) \right),
 \nonumber \\
g_3(m,r) = \frac{i}{2}\left(K_m(k^-_o r) - K_m(k^+_o r) \right).
\end{eqnarray}
Here $c_3$ and $d_3$ are arbitrary coefficients. At large values
of $r$ the functions $ f_3(m,r)$ and $ g_3(m,r)$ behave as
\begin{eqnarray}
f_3(m,r) &\sim & \sqrt{\frac{\pi}{2}}\frac{e^{-r \sqrt{v-e
-\beta^2/4}}}{(v-e)^{1/4} \sqrt{r}}
\cos\left(\frac{\beta r +\gamma}{2} \right) , \nonumber \\
g_3(m,r) &\sim & - \sqrt{\frac{\pi}{2}}\frac{e^{-r \sqrt{v-e
-\beta^2/4}}}{(v-e)^{1/4} \sqrt{r}}
 \sin\left(\frac{\beta
r + \gamma}{2}\right) ,
\end{eqnarray}
where $\gamma$ is defined as follows
\begin{equation}
\cos(\gamma)= \frac{\sqrt{v-e -\beta^2/4}}{\sqrt{v-e }} , \quad
\sin(\gamma) = \frac{\beta}{2 \sqrt{v-e }} .
\end{equation}
Thus, the radial wave functions $u_3(r)$ and $w_3(r)$ have the
appropriate behavior at infinity. It should be noted that the
radial wave functions have  the infinite number of zeros
 for the finite value of energy.

In the region 2 $(v=0)$ the radial equations may be written in the
suitable form
\begin{eqnarray}
r^2 \frac{d^2u}{dr^2} & + &r \frac{d u}{d r}  +( k^+_i k^-_i r^2 -
m^2)u
\nonumber \\
&= &( k^+_i -  k^-_i) r^2\left(\frac{d w}{d r} +  \frac{m+1}{r} w
\right) ,
\nonumber \\
 r^2\frac{d^2w}{dr^2} &+ &r \frac{d w}{d r}  +( k^+_i k^-_i r^2 -( m+1)^2)w
 \nonumber \\
&= &- ( k^+_i -  k^-_i) r^2 \left(\frac{d u}{d r}- \frac{m}{r} u
\right) ,
\end{eqnarray}
where
\begin{equation}
k^{\pm}_i(e,\beta) = \sqrt{e +\frac{\beta^2}{4}} \pm
\frac{\beta}{2} .
\end{equation}
Now in the case $\beta = 0$ ($k^+_i  = k^-_i$)  Eqs. (16) are the
Bessel equations.  Therefore we use the known properties
\cite{abr}
\begin{eqnarray}
 \left(\frac{d}{d r} + \frac{n}{r} \right) J_n(k r)
 &= &k J_{n-1}(k r) , \nonumber \\
 \left(\frac{d}{d r} - \frac{n}{r} \right) J_n(k r)
 &=& -k J_{n+1}(k r),
 \end{eqnarray}
\begin{eqnarray}
 \left(\frac{d}{d r} + \frac{n}{r} \right) Y_n(k r)
 &=& k Y_{n-1}(k r) , \nonumber \\
 \left(\frac{d}{d r} - \frac{n}{r} \right) Y_n(k r)
 &=& -k Y_{n+1}(k r)
 \end{eqnarray}
 of the Bessel functions  in order to obtain the exact solutions
\begin{eqnarray}
u_2(r) &= &c_{21} f_{21}(m,r) + d_{21} g_{21}(m,r)
\nonumber \\
&+ &c_{22} f_{22}(m,r) + d_{22} g_{22}(m,r),
\nonumber \\
 w_2(r) &=& c_{21} g_{21}(m+1,r) + d_{21} f_{21}(m+1,r)
 \nonumber \\
 &+ & c_{22} g_{22}(m+1,r) + d_{22} f_{22}(m+1,r)
\end{eqnarray}
 of system (16) where
\begin{eqnarray}
 f_{21}(m,r)& =& \frac{1}{2}\left(J_m(k^-_ir) + J_m(k^+_ir) \right),
 \nonumber \\
 g_{21}(m,r) &= &\frac{1}{2}\left(J_m(k^-_ir) - J_m(k^+_ir) \right),
\end{eqnarray}
\begin{eqnarray}
 f_{22}(m,r)& = &\frac{1}{2}\left(Y_m(k^-_ir) + Y_m(k^+_ir) \right),
 \nonumber \\
g_{22}(m,r) &=& \frac{1}{2}\left(Y_m(k^-_ir) - Y_m(k^+_ir)
\right).
\end{eqnarray}
Here $c_{21}, d_{21}, c_{22}$ and $d_{22}$ are arbitrary
coefficients.

The continuity conditions
 \begin{eqnarray}
u_1(r_i) &=& u_2(r_i), \quad
  u'_1(r_i)= u'_2(r_i), \nonumber \\
 w_1(r_i) &= &w_2(r_i), \quad
  w'_1(r_i)= w'_2(r_i),\nonumber \\
u_2(1) &=& u_3(1), \quad
  u'_2(1)= u'_3(1), \nonumber \\
 w_2(1) &=& w_3(1), \quad
  w'_2(1)= w'_3(1)
\end{eqnarray}
for the radial wave functions and their derivatives at the
boundary points $r=r_i$  and $r=1$ lead to the algebraic equations
\begin{equation}
M(m,e,v,\beta){\bf X} = 0
\end{equation}
for eight coefficients, where
\begin{equation}
{\bf X} =(c_1, d_1, c_{21}, d_{21}, c_{22}, d_{22}, c_3, d_3) ,
\end{equation}
and $ M(m,e,v,\beta)$ is $8 \times 8$ matrix
\begin{equation}
 M(m,e,v,\beta)=\left(\begin{array}{rr}
M_1 & M_2 \\
M_3 & M_4
 \end{array}\right),
 \end{equation}
 \begin{widetext}
 $$
 M_1=
  \left(\begin{array}{rrrr}
  f_1(m,r_i)&g_1(m,r_i)&
   -f_{21}(m,r_i)&-g_{21}(m,r_i) \\
 f'_1(m,r_i)&g'_1(m,r_i)&
   -f'_{21}(m,r_i)&-g'_{21}(m,r_i) \\
  -g_1(m+1,r_i)&f_1(m+1,r_i)&
   -g_{21}(m+1,r_i)&-f_{21}(m+1,r_i) \\
 -g'_1(m+1,r_i)&f'_1(m+1,r_i)&
 -g'_{21}(m+1,r_i)&-f'_{21}(m+1,r_i)
 \end{array}\right) ,
$$
$$
 M_2=
  \left(\begin{array}{rrrr}
  -f_{22}(m,r_i)&-g_{22}(m,r_i)&
   0&0 \\
 -f'_{22}(m,r_i)&-g'_{22}(m,r_i)&
   0&0 \\
  -g_{22}(m+1,r_i)&-f_{22}(m+1,r_i)&
   0&0 \\
 -g'_{22}(m+1,r_i)&-f'_{22}(m+1,r_i)&
   0&0
 \end{array}\right) ,
$$
$$
 M_3=
  \left(\begin{array}{rrrr}
  0&0&
   f_{21}(m,1)&g_{21}(m,1) \\
 0&0&
   f'_{21}(m,1)&g'_{21}(m,1) \\
  0&0&
   g_{21}(m+1,1)&f_{21}(m+1,1) \\
 0&0&
   g'_{21}(m+1,1)&f'_{21}(m+1,1)
 \end{array}\right) ,
 $$
 $$
 M_4=
  \left(\begin{array}{rrrr}
  f_{22}(m,1)&g_{22}(m,1)&
   -f_{3}(m,1)&-g_{3}(m,1) \\
  f'_{22}(m,1)&g'_{22}(m,1)&
   -f'_{3}(m,1)&-g'_{3}(m,1) \\
  g_{22}(m+1,1)&f_{22}(m+1,1)&
   -g_{3}(m+1,1)&f_{3}(m+1,1) \\
 g'_{22}(m+1,1)&f'_{22}(m+1,1)&
   -g'_{3}(m+1,1)&f'_{3}(m+1,1)
 \end{array}\right) .
$$
\end{widetext}
 Hence, the exact equation for energy spectrum is
\begin{equation}
\det M(m,e,v,\beta) =0. \end{equation}
 It should be stressed that in the explored model the number of
admissible energy levels is finite for the fixed angular momentum.
Note that Eq. (27) is invariant under two replacements $ m \to
-(m+1)$ and $\beta \rightarrow -\beta$.

If the energy values are found from Eq. (27) then it is simply to
get  the values of coefficients  $c_1,\, d_1,\, c_{21},\,
d_{21},\, c_{22},\, d_{22},\, c_3$ and $d_3$ from Eq. (24) and the
following normalization  condition
 $$
 \int_0^{\infty}\left(u^2(r) +  w^2(r)\right)r dr = 1 .
 $$

\section{Numerical and graphic illustrations}

Now we present some numerical and graphic illustrations in
addition to the analytic results.

Figures 1 and 2 demonstrate the  continuous radial wave functions
for  $m=1, e=17.88591, \beta=1, r_i = 0.2$ in the case $v=25$ and
for $m=1, e=21.4541, \beta=10,  r_i = 0.8$  in the case $v=100$,
respectively. Solid lines correspond to the functions $u(r)$ and
dashed lines correspond to the functions $w(r)$.

 Tables 1 and 2
show the dependence of energy $e$ on the ring width parameter
$r_i$  and the Rashba parameter $\beta$ for two values of angular
momentum number $m$   and two values of the well depth $v$. We see
that the number of energy levels increases  if the well depth $v$
increases.
\begin{table}
\caption{ Energy levels for $v=25$}\label{t1}
\begin{center}
\begin{tabular}{l r r r r  } \hline \hline
$r_i$&$\beta$&e &
\\ \hline
 \multicolumn{5}{c}{$m=0$}  \\ \hline
 $0.2$&$0$ &$5.58$  &$10.10$ &$22.69$  \\
 $0.2$&$1$ &$5.10$  &$10.07$ &$22.24$  \\
 $0.2$&$5$ &$-2.42$ &$3.61$  &$16.27$
  \\ \hline
 $0.5$&$0$ &$10.62$ &$13.09$ &         \\
 $0.5$&$1$ &$10.15$ &$13.05$ &         \\
 $0.5$&$5$ &$2.97$  &$7.39$  &
  \\ \hline
 $0.8$&$0$ &$19.90$  &$21.64$&         \\
 $0.8$&$1$ &$19.45$  &$21.59$&         \\
 $0.8$&$5$ &$12.82$  &$15.27$&
 \\ \hline
  \multicolumn{5}{c}{$m=1$}  \\ \hline
 $0.2$&$0$ &$10.10$  &$17.47$&         \\
 $0.2$&$1$ &$9.18$   &$17.86$&         \\
 $0.2$&$5$ &$-2.29$  &$9.59$ &
  \\ \hline
 $0.5$&$0$ &$13.09$ &$18.76$ &         \\
 $0.5$&$1$ &$12.18$ &$19.14$ &         \\
 $0.5$&$5$ &$1.37$  &$13.05$ &
  \\ \hline
 $0.8$&$0$ &$21.64$ &        &         \\
 $0.8$&$1$ &$20.78$ &        &         \\
 $0.8$&$5$ &$11.25$ &$18.18$ &
  \\ \hline \hline
\end{tabular}
\end{center}
\end{table}
 \begin{table}
\caption{ Energy levels for $v=100$}\label{t1}
\begin{center}
\begin{tabular}{l r r r r r r r } \hline \hline
$r_i$&$\beta$& & e &  & &
\\ \hline
 \multicolumn{8}{c}{$m=0$}  \\ \hline
 $0.2$&$0$ &$8.72$  &$12.77$ &$36.62$&$43.42$ &$79.52$ &$88.94$   \\
 $0.2$&$2$ &$6.97$  &$12.47$ &$34.95$&$43.08$ &$77.90$ &$88.50$   \\
 $0.2$&$10$&$-17.01$&$-14.43$&$8.23$ &$20.60$ &$53.15$ &$64.08$
  \\ \hline
 $0.5$&$0$ &$19.23$ &$21.22$ &$71.61$&$74.30$ &        &   \\
 $0.5$&$2$ &$17.54$ &$20.91$ &$69.98$&$73.91$ &        &   \\
 $0.5$&$10$&$-7.93$ &$-2.62$ &$45.67$&$50.05$ &        &
  \\ \hline
 $0.8$&$0$ &$54.30$  &$55.61$&       &        &        &   \\
 $0.8$&$2$ &$52.66$  &$55.24$&       &        &        &  \\
 $0.8$&$10$&$26.57$  &$32.94$&       &        &        &
 \\ \hline
  \multicolumn{8}{c}{$m=1$}  \\ \hline
 $0.2$&$0$ &$12.77$  &$21.74$ &$43.42$ &$57.91$  &$88.94$ &  \\
 $0.2$&$2$ &$9.56$   &$22.71$ &$40.43$ &$58.84$  &$86.19$ &  \\
 $0.2$&$10$&$-21.43$ &$-12.83$&$9.39$  &$36.34$  &$62.39$ &
  \\ \hline
 $0.5$&$0$ &$21.22$ &$27.03$&$74.30$&$81.32$  &  &  \\
 $0.5$&$2$ &$18.15$ &$28.05$&$71.39$&$82.15$  &  &  \\
 $0.5$&$10$&$-12.64$&$2.34$ &$45.78$&$59.81$  &  &
  \\ \hline
 $0.8$&$0$ &$55.61$&$59.49$& & &  &  \\
 $0.8$&$2$ &$52.69$&$60.38$& & &  &   \\
 $0.8$&$10$&$21.45$&$40.12$& & &  &
  \\ \hline \hline
\end{tabular}
\end{center}
\end{table}

\begin{figure}
\centering
\includegraphics{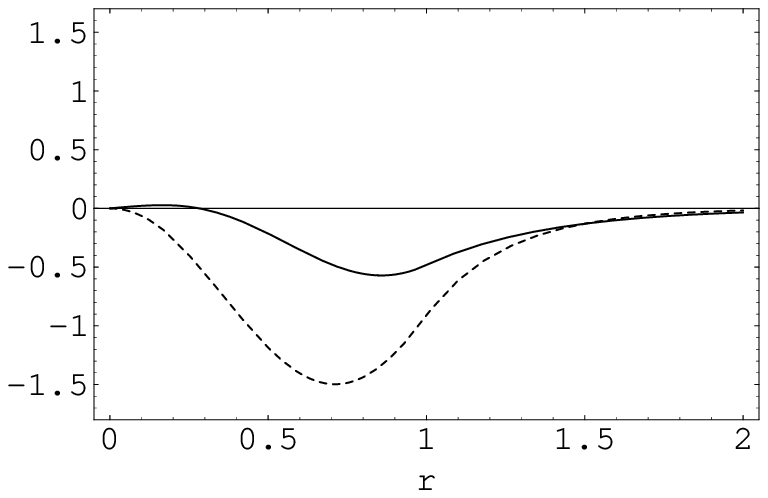} \caption{Radial wave functions  for $m=1, v=25, \beta=1, \\ r_i=0.2 $.
 Solid line for $u(r)$,  dashed line for $w(r)$. }\label{f1}
\end{figure}
 \begin{figure}
\centering
\includegraphics{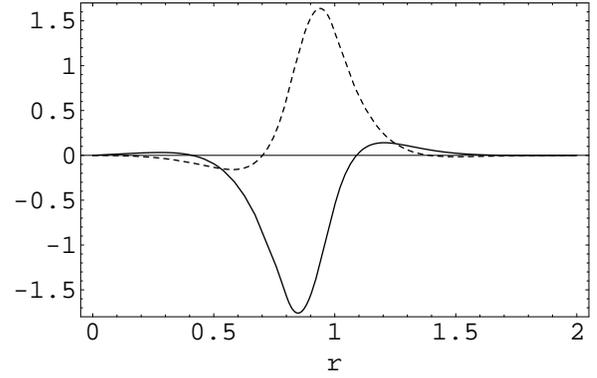} \caption{Radial wave functions  for $m=1, v=100, \beta=10, \\ r_i=0.8$.
 Solid line for $u(r)$,  dashed line for $w(r)$. }\label{f2}
\end{figure}

\section{Conclusion}

In our opinion the examined  exactly solvable  model  with the
realistic potential well of finite depth is physically adequate in
order to describe the behavior  of electron in a semiconductor
quantum ring of finite width taking into account the Rashba
spin-orbit interaction. Further we intend to generalize our
consideration by including the magnetic field effects.

\end{document}